 \documentclass[nolinenumbers]{aastex63}

\def\a{\"a}
\def\o{\"o}
\def\u{\"u}
\def\A{\"A}
\def\O{\"O}
\def\U{\"U}

\def\ang{\AA}

\def\gapprox{\lower.4ex\hbox{$\;\buildrel >\over{\scriptstyle\sim}\;$}}
\def\lapprox{\lower.4ex\hbox{$\;\buildrel <\over{\scriptstyle\sim}\;$}}
\def\ref#1{\par\noindent\hangindent1cm {#1}}

\begin{document}

\title{	   	Global Energetics in Solar Flares. XIII. 
		The Neupert Effect and Acceleration of Coronal Mass Ejections }

\correspondingauthor{Markus J. Aschwanden}

\author{Markus J. Aschwanden}
\email{aschwanden@lmsal.com}
\affiliation{Solar and Stellar Astrophysics Laboratory (LMSAL),
 Palo Alto, CA 94304, USA}

\begin{abstract}
Our major aim is a height-time model $r(t)$ of the propagation of 
{\sl Coronal Mass Ejections (CMEs)}, where the lower corona 
is self-consistently connected to the heliospheric path. 
We accomplish this task by using the Neupert effect to derive 
the peak time, duration,  
and rate of the CME acceleration phase, as obtained
from the time derivative of the {\sl soft X-ray (SXR)}
light curve. This novel approach offers the advantage
to obtain the kinematics of the CME height-time profile $r(t)$,
the CME velocity profile $v(t)=dr(t)/dt$, and the CME 
acceleration profile $a(t)=dv(t)/dt$ from {\sl Geostationary Orbiting
Earth Satellite (GOES)} and white-light data, without the need 
of {\sl hard X-ray (HXR)} data. We apply this technique
to a data set of 576 (GOES X and M-class) flare events observed
with GOES and the {\sl Large Angle Solar Coronagraph (LASCO)}. 
Our analysis yields acceleration rates in the range of
$a_A = 0.1-13$ km s$^{-2}$, acceleration durations of
$\tau_A = 1.2-45$ min, and acceleration distances in the
range of $d_A = 3-1063$ Mm, with a median of $d_A=39$ Mm,
which corresponds to the hydrostatic scale height of
a corona with a temperature of $T_e \approx 0.8$ MK. The results are
consistent with standard flare/CME models that predict 
magnetic reconnection and synchronized (primary) acceleration 
of CMEs in the low corona (at a height of $\lapprox 0.1\ R_{\odot}$), 
while secondary (weaker) acceleration may occur further out at 
heliospheric distances.
\end{abstract} 


\section{	Introduction			}

Eruptive processes, be it a geophysical volcano or a solar flare,
imply some causality between the triggering instability and 
secondary phenomena. The close connection between a solar flare
(observed in nonthermal hard X-ray (HXR) emission) and
the resulting coronal mass ejection (CME) (observed in white-light
and in extreme ultra-violet (EUV) dimming), 
is generally assumed to follow this causality order.
This two-step process may occur in near-simultaneous synchronization,
but delays between the two steps are 
caused by the processes of thermal heating and radiative cooling
(e.g., Aschwanden and Alexander 2001; Qiu 2021).
While the synchronized occurrence of eruptive flare and CME
events appears to be obvious, there are a lot of exceptions, such as 
flares without CMEs (in the case of non-eruptive or confined events),
or CMEs without flares (so-called stealth CMEs; Howard and Harrison 2013). 
Ambiguities in
the association of flares with CMEs occur frequently also, especially
when multiple flares occur during a single CME, while the opposite
case occurs less frequently. Another difficulty is insufficient
time resolution in the cadence of coronagraphs, amounting to $\approx 12$ min
for the {\sl Large Angle Solar Coronagraph (LASCO)} on board of
the {\sl Solar and Heliospheric Observatory (SOHO)}. 
Anyway, the close connection between solar flares and 
CMEs needs further investigation of these timing issues in a 
quantitative way.
The most novel aspect of this study is the application of the
Neupert effect, which predicts acceleration parameters and
allows us to test and model the flare/CME 
relationship from the lower corona out to the heliosphere,
even in the absence of HXR data.

The Neupert effect has been discovered in combined soft X-ray (SXR) 
and microwave emission (Neupert 1968; Kahler and Cliver 1988), and
has been dubbed the {\sl ``Neupert effect''} by Hudson (1991).
A close correlation between the HXR flux
from the {\sl Hard X-Ray Burst Spectrometer (HXRBS)}
onboard the {\sl Solar Maximum Mission (SMM)} and the 
time derivative peak of the 
{\sl Geostationary Orbiting Earth Satellite (GOES)} 
has been demonstrated by Dennis and Zarro (1993).
The physics of the Neupert effect has been modeled in terms of the
thick-target collisional bremsstrahlung process, which acts as the
main source of heating and mass supply (via chromospheric evaporation)
of the SXR-emitting hot coronal plasma (Veronig et al.~2005).
A {\sl ``theoretical Neupert effect''} was established by including 
variations in emission measure, temperature, radiative cooling losses,
conductive cooling losses, and low-energy cutoffs (Veronig et al.~2005;
Qiu 2021).

The temporal relationship between CMEs and associated solar flares,
observed with GOES and LASCO/SOHO, was found to co-evolve in a few flare/CME 
events (Zhang et al.~2001; Shanmugaraju et al.~2003; Zhang et al.~2004).
Larger statistical studies were performed with GOES and LASCO/SOHO data 
(up to 3217 CME events) (Moon et al.~2002, 2003), but no strong 
correlations were found. Significant correlations were found between
the CME kinetic energy and the GOES SXR flux (Burkepile et al.~2004),
or between the magnitude and duration of CME acceleration
(Zhang and Dere 2006; Vrsnak et al.~2007),
while other correlations between the duration of CME acceleration,
SXR rise time, and SXR peak flux (Maricic et al.~2007),
appear to be marginal.
Another study with 1114 flares, observed in HXRs with the
{\sl Burst and Transient Source Experiment (BATSE)}
onboard the {\sl Compton Gamma Ray Observatory (CGRO)} and 
SXRs (GOES), studied the relative timing of the SXR peak time 
and the HXR fluence, which was found to be consistent with the Neupert 
effect in about half of the cases (Veronig et al.~2002). 
It is suspected that the usage of single statistical quantities (such as 
the HXR end time, SXR peak time, HXR fluence, and HXR end time), 
neglects important information contained in the light curves 
(Veronig et al.~2002). However, flaring or non-flaring 
does not translate into two different types of CMEs (Vrsnak et al.~2005).
Tests of the Neupert effect using HXR data (from RHESSI) revealed
a synchronization within $\approx 5$ min between the hard X-ray emission 
(which is a direct indicator of the flare energy release) and the
CME acceleration profile (Temmer et al.~2008). More sophisticated
tests of the Neupert effect involve data from the
{\sl Extreme Ultra-Violet Imager (EUVI)} and coronagraph COR1 
onboard the {\sl Solar Terrestrial Relations Observatory (STEREO)},
which has a high cadence of $\lapprox 2.5$ min), and finds
CME acceleration at heights of $h \lapprox 0.4\ R_{\odot}$,
as well as CME velocity peaks at heights of $h \lapprox 2.1\ R_{\odot}$
(Temmer et al.~2010).

In this paper we make for the first time use of the Neupert effect to constrain
the height-time profile of CMEs propagation in the lower corona, 
which bridges to the CME trajectory in the heliosphere in a self-consistent
way, using GOES and LASCO/SOHO data. We describe the methodology in
Section 2, the data analysis and results in Section 3, discussions 
in Section 4, and conclusions in Section 5. 

\section{	Methodology 			}

This study builds on three previous works on the kinematics
and energetics of CMEs, which are based on 
the EUV dimming method (Aschwanden 2016; Paper IV), 
self-similar adiabatic expansion (Aschwanden 2017; Paper VI),
and the aerodynamic drag force (Cargill 2004; 
Vrsnak et al.~2004, 2013; Aschwanden and Gopalswamy 2019; 
Paper VII), while the present study includes the Neupert effect. 

\subsection{	The Neupert Effect			}

The Neupert effect was first pointed out by Werner Neupert (1968),
who noticed that time-integrated microwave fluxes closely match the
rising portions of SXR emission curves, which has been demonstrated 
by others also (e.g., Kahler and Cliver 1988), and was later
dubbed the {\it Neupert effect} by Hudson (1991). Estimating the
time-integrated microwave fluxes from the nonthermal HXR 
emission, the Neupert effect has been generalized, where the time 
derivative of the SXR, i.e., $dF_{\rm SXR}(t)/dt$, 
can be considered as a proxy for the 
HXR emission $F_{\rm HXR}(t)$ (Dennis and Zarro 1993; 
Veronig et al.~2002),
\begin{equation}
	F_{\rm HXR}(t) \propto {dF_{\rm SXR}(t) \over dt } \ .
\end{equation}
The SXR flux $F_{\rm SXR}(t)$ can most conveniently be
obtained from GOES 1-8 \ang\ data, which is generally available
during the last 46 years (given in units of $W m^{-2}$). 
Using the Neupert effect, we can define 
the peak flux $F_{\rm peak,HXR}$ of the HXR emission,
\begin{equation}
	F_{{\rm peak,HXR}} = F_{\rm HXR}(t=t_{\rm peak,HXR}) = max[F_{\rm HXR}(t)] 
	\qquad t_{\rm start,GOES} \le t \le t_{\rm peak,GOES} \ ,
\end{equation}
within the time interval $[t_{\rm start,GOES}, t_{\rm peak,GOES}]$.
The time profile of a GOES flare is characterized
by the start time $t_{\rm start,GOES}$, the peak time $t_{\rm peak,GOES}$, 
and the end time $t_{\rm end,GOES}$.

The novel application of the Neupert effect here is the model assumption
that the start time $t_{\rm start,CME}$ of CME acceleration can be equated 
to the start time $t_{\rm start,HXR}$ of the HXR flux,
which is calculated from the time derivative of the GOES SXR flux
(Eq.~1),
\begin{equation}
	t_{\rm start,CME} \approx t_{\rm start,HXR} \ .
\end{equation}
This assumption can be justified because the start time of a HXR
time profile (or light curve) marks the start time of the energy release
during a (CME-associated) flare, which is also the start time of the 
force that launches a CME. While this assertion (Eq.~3) appears to be
novel and untested, we will see that the start time $t_{\rm start,GOES}$ 
of the Neupert-constrained relationship matches the CME start time 
$t_{\rm start,CME}$ (Eq.~3) in most of the events.

We are using the high time resolution GOES data, which are binned in
intervals of $\Delta t = 3\ s$. However, in order to have a smooth
time profile we average them with a boxcar length of 60 time bins, 
which corresponds to an effective time resolution of $\Delta t_{smooth} = 3$ min, 
and is sufficient to resolve the shortest flare durations. This implies that
the Neupert-constrained flare peak times are also subject to the same
accuracy of $\Delta t \approx 3$ min.

\subsection{	Self-Similar Expansion of CMEs		}

First we determine the starting height $r_0=r(t=t_0)$ of a CME
at the peak time $t_0$ of the SXR flux derivative, which is a proxy
for the HXR peak time according to the Neupert effect (Eq.~2).
We assume a model of self-similar spherical expansion, where the
center of the spherical CME bubble starts with a point-like
geometry at the start time at a photospheric 
distance $r=R_{\odot}$ from Sun center. It does not matter if we assume
a photospheric or coronal level, since the height difference
between the bottom and top of the chromosphere 
is only about 0.3\% of a solar radius $R_{\odot}$. During the 
spherical expansion, the CME (leading edge) front has a self-similar 
geometry in all directions, along the plane-of-sky, as well as in any 
other direction. The location of the CME start can be specified 
with heliographic longitude and latitude $(l, b)$, and the projected 
distance of the CME with respect to Sun center at the peak time $t_0$ 
amounts to,
\begin{equation}
	r_0 = r(t=t_0) = R_{\odot}\ \sin{ \sqrt{(l^2+b^2)}} \ ,
\end{equation}
which has the minimum limit of $r_0=0$ for a halo CME originating at disk
center, and $r_0 = 1.0\ R_{\odot}$ for a CME starting at the solar limb.

In the present study we are using LASCO/SOHO data, but the same procedure
can be applied to other full-disk white-light data with heliospheric coverage. 
The LASCO height-time
profiles consist of a time series of height-time measurements $r_i=r(t_i)$,
$i=1,...,n$, where the first measurement is taken at a mean altitude of
$r_1 \approx 3 R_{\odot}$, and the last is taken at $r_n \approx 30\
R_{\odot}$ (see also Fig.~5 in Paper VII). From the LASCO data archive, only
the coronagraphs C2 and C3 data have been used for uniformity, because
LASCO/C1 was disabled in June 1998.

The Neupert effect predicts the start time $t_0$ of the CME 
from the time derivative of the GOES SXR profile (according to 
Eq.~3), while the (projected) starting distance $r_0=r(t=t_0)$ is given
by the heliographic position (Eq.~4), so that we can add this
initial data point $[t_0, r_0]$ to the height-time observations
of the LASCO white-light time series $t_{obs,i}, i=1,...,n$.

\subsection{Kinematic Model of CME Acceleration 	}

The height-time profile $r(t)$ of a CME can in the simplest
way be characterized by an initial acceleration phase
(during the time interval $t_A \le t \le t_B$) and a 
subsequent expansion with constant velocity (during the
time interval $t \ge t_B$), which we define 
in terms of the acceleration rate profile $a(t)$
being a step-function,
\begin{equation}
	a(t) = {dv(t) \over dt} = \left\{
	\begin{array}{ll}
        a_A   & {\rm for\ } t_A \le t \le t_B \\
	0     & {\rm for\ } t \ge t_B \\
	\end{array} 
	\right. \ .
\end{equation}
By integrating the acceleration time profile $a(t)$ we
obtain the velocity profile $v(t)$, 
\begin{equation}
	v(t) = \int_{t_A}^t a(t) dt = \left\{
	\begin{array}{ll}
	a_A (t-t_A)    & {\rm for\ } t_A \le t \le t_B \\
	v_B            & {\rm for\ } t \ge t_B \\
	\end{array} 
	\right. \ ,
\end{equation}
which essentially yields a linear increase of the CME
velocity $v(t)$ during the acceleration phase, and a
constant velocity $v(t)$ after the end time of the
acceleration phase, with $v(t)=v_B$ for $t \le t_B$.
Now we can calculate the time
integral to obtain the height-time profile, $r(t)$,
\begin{equation}
	r(t) = \int_{t_A}^t v(t)\ dt = \left\{
	\begin{array}{ll}
	r_A + {1 \over 2} a_A (t-t_A)^2   & {\rm for\ } t_A \le t \le t_B \\
	r_1 + v_B*(t-t_1)                 & {\rm for\ } t \ge t_B \\
	\end{array} 
	\right. \ .
\end{equation}
No linear term $v_A (t-t_A)$ is included because it is more
realistic to start acceleration from rest ($v_A = 0$).
The distance $r_B$ at the end time $t = t_B$ of the acceleration
phase follows then from Eq.~(7),
\begin{equation}
	r_B = r_A + {1\over 2} a_A (t_B-t_A)^2 \ ,
\end{equation}
and the CME velocity $v_B = v(t=t_B)$ follows from Eq.~(6),
\begin{equation}
	v_B = a_A (t_B-t_A) \ .
\end{equation}
This final CME velocity $v_B$ after the end of the acceleration phase
can then be obtained from the white-light data 
$[r_1, r_2, ..., r_i, ..., r_n]$ at the times
$[t_1, t_2, ..., t_i, ..., t_n]$,
\begin{equation}
	v_B = {(r_i - r_1) \over (t_i - t_1)} \ ,
\end{equation}
where the variable $r_i = r(t=t_i)$ can be arbitrarily
selected among the observables $i=2,...,i,...,n$.
In principle one could choose a linear regression method,
or a polynomial fit, 
but our tests indicate that the choice of the fitting range
leads to larger uncertainties than the formal error of the linear
regression or polynomial fit method. Here we choose
a value of $i=5$, which is a suitable compromise between
data noise reduction (which affects the extrapolation most for $i=2$) 
and the detection of nonlinear trends (which affects the extrapolation
most for $i=n$).

From the expression of $v_B$ (Eq.~9) we obtain the 
acceleration parameter $a_A$,
\begin{equation}
	a_A = {v_B \over (t_B-t_A)} = {v_B \over \tau_A} \ ,
\end{equation}
from which the other parameters can be derived,
such as $r_B$ (Eq.~8) and $v_B$ (Eq.~9).

The Neupert model provides the peak time $t_0$ of 
acceleration and the duration of acceleration,
$\tau_A=(t_B-t_A)$, which defines the start time
$(t_A)$ and end time $(t_B)$ of the acceleration phase,
\begin{equation}
	t_A = t_0 - {\tau_A \over 2} \ ,
\end{equation}
\begin{equation}
	t_B = t_0 + {\tau_A \over 2} \ .
\end{equation}

Let us summarize the physical parameters of this CME
height-time model that consists of an acceleration phase
($t_A \le t \le t_B$) and a propagation phase
($t_B \le t \le t_n$), where only the time range
of ($t_1 \le t \le t_n$) can be fitted, depending on the
range of available white-light data. 
Thus we have 6 time markers ($t_A, t_0, t_B, t_1, t_i, t_n$),
6 distances from Sun center ($r_A, r_0, r_B, r_1, r_i, r_n$),
one CME velocity ($v_B$), and one acceleration rate ($a_A$).
The observed data variables are presented in Figs.~(1-6)
and Tables (1-3), the theoretical model is shown in Fig.~(7), and
the analyzed distribution functions and correlations are shown
in Figs.~(8-9).

\section{	Data Analysis and Results		}

\subsection{	Observations 				}

In the previous studies (Papers IV and VI) we analyzed all 399
GOES $\ge$ M1.0 class flare events observed during the first 3.5 
years of the SDO mission (2010 June 1 - 2014 January 31),
which we expanded to 576 events (2010 June 1 - 2014 November 16)
in Paper VII and in this study here. From the LASCO data we 
extract height-time profiles that encompass the GOES 1-8 \ang\  
flare durations. The LASCO/SOHO catalog
(https://cdaw.gsfc.nasa.gov/CME$\_$list) is based on visually 
selected CME events, and is created and maintained by Seiji Yashiro
and Nat Gopalswamy (Yashiro et al.~2008; Gopalswamy et
al.~2009, 2010). The LASCO catalog provides height-time
profiles $r(t)$ with a typical cadence of 
$\Delta t_{\rm LASCO} t \approx 12$ min,
while we smoothed the GOES time profiles with a box car of
$\Delta t_{\rm GOES} = 3$ min.
The entire data analysis is carried out with an automated
CME detection algorithm without any human interaction.

\subsection{	Examples of Data Analysis		}

From the total sample of 576 analyzed events we select 
six special categories of CME events that are presented
in Figs.~1-6 and in Table 1, rendering four examples in each group.
The full dataset of 576 analyzed events is tabulated
in Table 2, available as a machine-readable file.

\underbar{Group A: Largest (GOES X-class) flares (Fig.~1):} 
This first group of four events represents the largest
GOES flares, which range from GOES class X3.3 to X6.9
for the four events shown in Fig.~1.
The largest event \# 61 (Fig.~1a) has a GOES class X6.9,
exhibiting a single-peaked GOES light curve above a
background level of GOES C class (black curve in 
Fig.~1a), and the time derivative of the GOES light curve shows
a single-peaked curve too (red curve in Fig.~1a),
which defines the Neupert-constrained CME peak time 
$t_0$ (vertical red line).
The GOES flare (indicated with a hatched area in Fig.~1a), 
as cataloged by NOAA, has a start time
of 2011-08-09 07:48 UT, a peak time of 08:05 UT,
and an end time of 08:08 UT, which defines the flare duration,
$\tau_{\rm flare}=(t_{\rm end}-t_{\rm start})=20$ min. 
The times are given here 
in units of hours relative to the midnight of the corresponding day.
The LASCO detection time range (as indicated with 
vertical dotted lines), is $t_1=t_{\rm start,LASCO}=8.20$ hr, 
and $t_n=t_{\rm end,LASCO}=10.70$ hr.
The initial location of the CME is at a distance of 
$r_0=0.95\ R_{\odot}$, based on the flare location with 
heliographic position N20W69 (Eq.~4). 
The start time $t_0$ of the CME is obtained
from the peak of the time derivative of the
GOES light curve, evaluated in the time interval between
$t_{\rm GOES,start}$ and $t_{\rm GOES,peak}$, following the  
rule of the Neupert effect. We also measure the duration
of the CME acceleration from the full width at half maximum
(FWHM) of the time derivative
peak, which amounts here to $\tau_A = 2.4$ min.
Note that the duration of CME acceleration ($\tau_A=2.4$ min)
is much shorter than the GOES flare duration ($\tau_{\rm flare}=20$ min) 
(as defined by NOAA).
The final velocity at $t=t_n$ of the CME amounts to 
$v_B=1830$ km s$^{-1}$, and the acceleration rate is
$a_A=12.8$ km s$^{-2}$.

The next largest event is of GOES class X5.4 (Fig.~1b),
which exhibits an 
acceleration duration of $\tau_A = 5.8$ min,
a final CME velocity of $v_B=2813$ km s$^{-1}$,
and an acceleration rate of $a_A=8.1$ km s$^{-2}$.
There are actually four peaks visible in the time
derivative of the SXR flux, but since the main peak
coincides closely with the first detection time of
the CME, we consider this CME event as unambiguously
associated with the near-simultaneous GOES flare. 

The third event (Fig.~1c) shows a relatively late
detection of the CME after a delay time of 
$(t_1-t_A)=1.507-0.740=0.77$ hr) and at a distance of 
$d_A=(r_1-r_A)=11.170-0.992=10.2\ R_{\odot}$.
This example demonstrates that our linear extrapolation
scheme is fairly robust in determining the CME start time 
that is assumed to coincide with the start of the
flare acceleration time $t_A$, even when the CME detection 
occurs late.
 
The fourth event is of GOES class X3.3 (Fig.~1d) and
indicates a deceleration of the CME velocity $v(t)$
during the detection of white-light data (in the entire
time interval from $[t_1,t_n]$), while the range $[t_1,t_5]$ 
that we use in our extrapolation scheme (Eq.~10) indicates an 
initially steeper slope, and thus implies a higher velocity.  
Thus, this example illustrates how the accuracy of the
CME velocity $v_B$ depends on the choice of the fitting
range.

\medskip
\underbar{Group B: Smallest GOES M-class flares (Fig.~2):} 
The four smallest GOES M-class flares reach systematically
lower altitudes than the large X-class flares
during the LASCO detection time window, i.e., 
$r_2 \approx 5-10\ R_{\odot}$, compared with
$r_2 \approx 30\ R_{\odot}$ for large X-class flares 
(e.g., Fig.~1).
The four examples shown in Fig.~2 all display a small
CME velocity of $v_B \approx 200-500$ km s$^{-1}$,
which appears to be typical for weak GOES class flares
(of $\approx$ M1 class). 

Examining the extrapolated height-time profiles $r(t)$
in Fig.~2 we notice that they exhibit various degrees of
initial ``jumps'' at the beginning of their profile $r(t)$,
from small jumps (event \#293; Fig.~2d) to large jumps
(event \#221; Fig.~2b). Such jumps could occur due to
four possible reasons: (i) The initial expansion is much
faster than later on during the (heliospheric) expansion
(similar to the cosmological inflationary model);
(ii) Stealth CMEs that start at an altitude of
$r \approx 3-5\ R_{\odot}$; 
(iii) The CME-associated flare started earlier than
identified with our automated detection method, 
which searches within a finite time window of  
$[t_0-2.0, t_0+0.5]$ hr; 
(iv) Erroneous detection or confusion of white-light
CME observations, especially for asymmetric halo CMEs.
The inflationary scenario (i) would require two different
driver mechanisms or a rapid change in the expansion rate.
A stealth-CME scenario (ii) requires an invisible driver,
and options (iii) and (iv) imply unlikely large errors
in the reported height-time plots of LASCO, so it is
not obvious how to explain this phenomenon.

\medskip
\underbar{Group C: Largest CME (detection) distances (Fig.~3):} 
The largest CME detection distances in LASCO data amount
to $r_n \approx 30\ R_{\odot}$. 
Since the flare event \#146 (Fig.~3a) and \#147 (Fig.~3b) occur 
only $\approx 0.2$ hr apart, they cannot be properly disentangled,
but it is possible that two CMEs occur in rapid succession.
The height-time plot of this two events is close to linear,
which implies an almost constant CME velocity.
A similar situation occurs for event 
$\#406$ (Fig.~3c) and $\#407$ (Fig.~3d),
where one single CME is reported (from LASCO data), while
multiple GOES flares occur during the same CME event,
which demonstrates possible ambiguities in the association
of flares with CMEs. 
The latter two events exhibit a very low velocity,
$v_B=195$ km s$^{-1}$, which seems to accelerate after the
initial acceleration phase, most likely indicating 
a second acceleration phase in the solar wind.

\medskip
\underbar{Group D: Longest CME durations (Fig.~4):} 
Case 4d reveals double CMEs where one overtakes 
a previous CME, similar to the ``Cannibalistic CMEs''
reported by Gopalswamy et al.~(2001).
Figs.~(4a), (4b), and (4c) all show cases with multiple GOES 
SXR peaks during a single CME detected in LASCO,
indicating that matching of GOES flares with CME events 
can be ambiguous. Moreover, all four events with the
longest CME duration exhibit very low initial CME 
velocities ($v_B=41-104$ km s$^{-2}$), similar to
the CMEs associated with the smallest GOES flares
(Fig.~2), and thus may be subject of the same 
scenarios discussed in group B.

\medskip
\underbar{Group E: Fastest CME velocities (Fig.~5):} 
These fastest CMEs have velocities of $v_B \approx 2200-3000 $
km s$^{-1}$ and are associated with large flares
(of GOES class M8.7 to X4.9). All four events shown
in Fig.~5 have similar characteristics: high velocities
and high acceleration rates ($a_A \approx 4-10$ km s$^{-2}$).

\medskip
\underbar{Group F: Slowest CME velocities (Fig.~6):} 
These slowest CMEs have initial velocities of $v_0 \approx 130-230 $
km s$^{-1}$ and are associated with the weakest analyzed
GOES flare events, from M1.0 to M1.3 class. 
However, the NOAA flare time range (hatched areas
in Fig.~6) does not coincide with the Neupert peak time
identified near the LASCO-detected CME start time,
which indicates some ambiguity in the association
of CME/flare events.  

\medskip
In summary, these 24 cases presented in Figs.~1-6
show a wide variety of cases and illustrate both
success and problems of connecting height-time plots
from the corona to the heliosphere.
The described 24 events can be considered as a 
representative sample out of the 576 analyzed flare events. 
Problems occur when 
(i) a single CME is detected during multiple GOES flares
which leads to ambiguities in the association of flares to CMEs,
(ii) when the first white-light measurement of the CME front
occurs before the Neupert peak time; or
(iii) when an initial acceleration and deceleration phase 
is temporally not resolved.
Nevertheless, besides these few problematic cases,
the Neupert model was found to be adequate in most cases.
We present some statistical information in the next section.

\subsection{	Flare/CME Event Statistics 	}

From the total analyzed data set of 576 flare/CME events, 
a subset of 373 events ($65\%$) have a physical solution 
that is consistent with the Neupert-constrained model 
of the flare/CME timing (Table 3). 

A fraction of 131 flare/CME events
($23\%$) violates the Neupert rule that the start of the
flare phase preceeds the start of the CME acceleration
phase, although the time difference often amounts
to the time resolution of LASCO data ($\Delta t = 3$ min).
If we correct those events by eliminating the first time bin,
the number of Neupert-consistent flare/CME events improves 
by 47 events ($8\%$) (Table 3). 

Another problem in automated CME detection is the
ambiguity between CMEs and flares, affecting 84 cases
($15\%$) in our study. Most of these cases show multiple 
flare events during a single CME, while the opposite 
case is rarely observed. Flares without CMEs can occur, 
especially for non-eruptive or confined flares. 

\subsection{	Statistics of CME Variables	}

In Fig.~(8) we show the size distributions of (logarithmic)
CME variables, including the GOES flux $F_{\sl SXR}$ (Fig.~8a),
the GOES rise time $t_{rise}=t_{peak,GOES}-t_{start,GOES}$ (Fig.~8b),
the duration of CME acceleration $\tau_A=(t_B-t_A)$ (Fig.~8c),
the CME velocity $v_B$ (Fig.~8d),
the CME acceleration distance $d_A = (r_B-r_A)$ (Fig.~8e), and
the CME acceleration rate $a_A$ (Fig.~8f).
The minimum, median, and maximum values of each distribution
is listed in each plot.
The distributions of the GOES flux and the acceleration time
follow approximately an exponential function, which indicates
a random process, while all other distributions follow
approximately a (log-normal) Gaussian-like function. 
The median values can be considered as ``typical values''.
The median GOES class is M2.0.
The SXR rise time is $\tau_{rise}=9.0$ min, which is a factor of 
three longer than the CME acceleration time $\tau_A=3.0$ min.
The mean CME velocity is $v_B = 386$ km s$^{-1}$,
which agrees with the slow solar wind speed.
The median acceleration distance is $d_A=39$ Mm, which
corresponds to the electron density scale height
$\lambda \approx 50$ Mm ($T_e$ [MK]), fitting a coronal
temperature of $T_e \approx 0.8$ MK. 
The median CME acceleration rate is $a_A=2.0$ km s$^{-2}$,
but varies in the range of $a_A=0.1-13.5$ km s$^{-2}$. 

\section{	Discussion				}

The close connection between solar flares and CMEs 
has been studied in many different wavelengths, 
especially in SXRs and HXRs for flares,
and in white-light and EUV for CMEs. Quantitative
relationships are generally expressed by flux time
profiles $F(t)$ for flares, and by kinematic time
profiles for CMEs, such as height-time profiles $r(t)$,
velocity profiles $v(t)=dr(t)/dt)$, and acceleration
profiles $a(t)=dv(t)/dt$. The most commonly used
timing parameters are the start times $t_{start}$,
the peak times $t_{peak}$, the rise times
$\tau_{rise}=(t_{peak}-t_{start}$), the CME acceleration 
time duration $\tau_A=(t_B-t_A)$, and the flare duration 
$\tau_{\rm flare}=(t_{\rm end}-t_{\rm start})$.
We discuss the most relevant studies and findings 
in the following. 

\subsection{	CME Acceleration and Soft X-rays 	}

A close correlation between CME acceleration and solar flare
SXR emission has been found and corroborated over 
the last two decades. Analyzing CME events with 
LASCO/SOHO and EIT/SOHO data, correlations were found between 
the time evolution of the CME velocity $v_{\rm CME}(t)$ and 
the time evolution of the GOES SXR flux 
$F_{\rm SXR}(t)$ (Zhang et al.~2001, 2004; 
Shanmugaraju et al.~2003; Maricic et al.~2004, 2007;
Burkepile et al.~2004; Vrsnac et al.~2007),
\begin{equation}
	v_{\rm CME}(t) \propto F_{\rm SXR}(t) \ .
\end{equation}
A similar time evolution between the CME velocity
and SXR flux implies also similar peak times
in the two types of emission, and perhaps a scaling
law between the peak values of the two emissions.
Such a scaling law has been suggested from measurements
of the two quantities at their peak times $t_{\rm peak}$
(Moon et al.~2002, 2003; Maricic et al.~2007; 
Vrsnac et al.~2004, 2005, 2007; Zhang and Dere 2006), 
\begin{equation}
	v_{\rm peak,CME} \propto F_{\rm peak,SXR} \ ,
\end{equation}
for which we find a marginal correlation only,
e.g., CCC$\approx 0.41$ only (Fig.~9a). 

Since the acceleration $a(t)=dv(t)/dt$ is by definition 
the time derivative of the velocity profile $v(t)$,
we also expect the following relationship 
between the CME acceleration $a_{\rm CME}(t)$
and the GOES time derivative $dv_{\rm CME}(t)/dt$,
\begin{equation}
	a_{\rm CME}(t) \propto {dv_{\rm CME}(t) \over dt}
	\propto {dF_{\rm SXR}(t) \over dt} 
	\propto F_{\rm HXR} \ .
\end{equation}
This relationship is related to the Neupert effect (Eq.~1),
if the time derivative of the SXR time profile
$dF_{\rm SXR}/dt$ is taken as a proxy for the HXR
light curve $F_{\rm HXR}(t)$ and the acceleration rate
$\alpha_{\rm CME}(t)$. The correlation between the
acceleration rate $a_A$ and the GOES flux is shown in Fig.~(9c),
which shows a marginal cross-correlation coefficient of
CCC=0.29 and a linear regression fit with a slope of
0.29 (Fig.~9c). The only strong correlations are found 
between the CME velocity $(v_B)$ and the CME
acceleration distance $d_A$ (CCC=0.73; Fig.~9d),
and the acceleration distance $d_A$ and the CME
acceleration duration $\tau_A$ (CCC=0.70; Fig.~9b).
These strong correlations can be explained by the
kinematic relationships $v_B = a_A \tau_A $ (Eq.~9)
and $v_B = d_A / \tau_A$.

A so-called ``indirect flare-proxy method'' has been
defined to characterize the temporal relationship
between CMEs and flares, based on the assumptions that
(i) the rise time of the associated SXR flare
equals the CME acceleration time $\tau_{\rm A_a}$ 
(Zhang and Dere 2006),
\begin{equation}
	\tau_{\rm rise,SXR} = \tau_A \ ,
\end{equation}
and (ii) the average velocity in the outer corona 
equals the velocity increase during the acceleration 
phase (Zhang and Dere 2006). However, here we find
that the SXR rise time $\tau_{\rm rise,SXR}$
is not correlated with the CME acceleration duration
$\tau_A$ (Fig.~9e). The SXR rise time appears to include
preflare activities that are not related to the HXR
impulsive phase, which is consistent with the fact
that the mean SXR rise time, i.e., ($\tau_{\rm rise,SXR}=9.0$
min (Fig.~8b), is a factor of $\approx 3$ longer
than the mean acceleration duration ($\tau_A=3.0$
min) (Fig.~8c).

Since not all CMEs are accompanied by solar flares,
it was investigated whether flare-associated CMEs and
flare-less CMEs have different physical parameters,
but it turned out that both data sets show quite 
similar characteristics, contradicting the concept 
of two distinct (flare and flare-less) types of CMEs. 
(Vrsnak et al.~2005).

\subsection{	CME Acceleration and Hard X-Rays   	}

Now we shift our discussion from SXRs to HXRs.
A close synchronization between the CME acceleration profile
and the flare energy release, as indicated by the RHESSI HXR
flux, was found in several CME events, where the HXR peak
time and the CME acceleration start occurs within minutes
(Temmer et al.~2008; 2010).

Temmer et al.~(2008) analyzed the relationship between fast halo
CMEs and the synchronized flare HXR bursts for two events,
i.e., the X3.8 GOES class event on 2005 January 17, and the
M2.5 GOES class event on 2006 July 6, using HXR data
from RHESSI and white-light data from LASCO/SOHO. The HXR 
energy ranges are $\gapprox 50$ keV, and $\gapprox 30$ keV,
respectively. The height-time plots of the distance $r(t)$
from the Sun center, as well as the CME velocity profiles
$v(t)$ and acceleration profiles $a_{\rm HXR}(t)$ were derived, which
clearly demonstrate that the HXR light curve $F_{\rm HXR}(t)$
from RHESSI is highly correlated with the acceleration $a_{\rm HXR}$ 
of the CME, as measured from the time derivative of the CME
height-time plot, synchronized within $\approx 2$ to 5
minutes.

Another three events (2007 June 3, C5.3; 2007 December 31, C8.3;
2008 March 25, M1.7) were analyzed and compared with numerical
simulations in Temmer et al.~(2010). The distance $r(t)$ of 
the CME leading edge was measured from STEREO A data, and
HXR time profiles $F_{\rm HXR}(t)$ from $>50$ keV
RHESSI data. The usage of STEREO data provided coverage of
the coronal range of $r \lapprox 3.0\ R_{\odot}$, which is
not available in LASCO data (which is also the case for all
LASCO events analyzed here). The improved data analysis method
yielded relatively small time differences of $\Delta t$= 
0.1, 2.0, and 1.5 minutes between the acceleration peak time
and the HXR peak time.

The purpose of this study has a very similar goal as the two
previous studies of Temmer et al.~(2008; 2010), namely the
establishment of the time coincidence between solar flare
HXR start times and CME acceleration start times.
However, instead of using the HXR data from RHESSI
(which has been decommissioned on 2018 August 9), we are
using GOES 1-8 \ang\ data here and apply the Neupert effect,
which yields a fairly reliable proxy for the HXR 
peak time as calculated from the time derivative
of the SXR time profiles (e.g., from GOES).
Our strategy is to extrapolate the height-time profiles
of LASCO-observed CMEs to the initial coronal height,
which supposedly coincides with the Neupert-predicted
HXR timing. Temmer et al.~(2008; 2010)
claim an accuracy of a few minutes in the relative timing,
which corresponds to the typical time resolution of
$\Delta t_{\rm LASCO} \approx 3$ min used here. 
We expect that the time resolution constitutes an upper
limit on the observed time scales. Indeed, by
measuring the full widhts of half maximum (FWHM)
of the time derivatives in the GOES flux time
profiles, we obtain a compatible mean value of 
$\tau_A = 3.0$ min; (Fig.~8c).

\subsection{	CME Acceleration Parameters		}

Our Neupert-constrained model provides four independent parameters
of the acceleration mechanism for each flare/CME event, based on 
the measurement of the peak time $t_0=(t_A+t_B)/2$ and the acceleration duration 
$\tau_A=(t_B-t_A)$. 
This includes the start time $t_A$ (Eq.~12), the end time $t_B$ of the CME 
acceleration phase (Eq.~13), the mean acceleration rate $a_A$ (Eq.~11, Fig.~8f),
and the CME acceleration distance $d_A=(r_B-r_A)$, (Fig.~8e). The distributions,
median values, and ranges are given in Fig.~(8), where we find, 
\begin{equation}
	\tau_{A,med} = 3.0\ {\rm [min]} \ , \quad 1.2 < \tau_A < 44.5 \ [{\rm min}] \ , 
\end{equation}
\begin{equation}
	a_{A,med} = 2.0\ {\rm [km\ s^{-2}]} \ , \quad 0.1 < a_A < 13.5 \ [{\rm km}\ s^{-2}] \ , 
\end{equation}
\begin{equation}
	d_{A,med} = 39\ {\rm [Mm]} \ , \quad 2.8 < d_A < 1063 \ [{\rm Mm}] \ . 
\end{equation}
Compatible acceleration rates were measured in other data sets: $a_A=3$ km s$^{-2}$
in an X1.6 flare, and $a_A=0.2-0.4$ km s$^{-2}$ in a M1.0 flare (Qiu et al.~2004).
The determination of such parameters in the acceleration of CMEs,
entirely based on the Neupert effect, are obtained with unprecedented statistics
here.  

\subsection{	Primary CME Acceleration in Hydrostatic Corona 	}

What new insights does the Neupert model convey? A key result is the
spatial location of the CME acceleration, which we find to be confined 
within a median distance of $d_{A,med}=39$ Mm above the solar photosphere.
Incidentally, this spatial scale corresponds closely to the hydrostatic
scale height $\lambda$ of the corona in the Quiet-Sun and in coronal hole regions,
having a typical value of $\lambda_T \approx 47$ Mm for a coronal
temperature of $T_e \approx 0.8$ MK (e.g., Aschwanden 2004, p.69),
\begin{equation}
	\lambda(T_e) = {2 k_B T_e \over \mu m_H g_{\odot}}
		\approx 47 \left( {T_e \over 1\ {\rm MK}} \right) 
		\quad {\rm MK} \ ,
\end{equation}
where $k_B$ is the Boltzmann constant, $\mu \approx 1.27$ is the
mean molecular weight (for a H:He=10:1 ratio), $m_H$ is the mass
of a hydrogen atom, and $g_{\odot}=2.74 \times 10^4$ cm s$^{-2}$
is the solar gravitation. Thus, our statistical study is consistent
with an acceleration height located in the lowest electron density
scale height of the hydrostatic corona. If the CME
propagates along a streamer, the mean coronal density and
temperature $T_e$ can easily vary by about a factor of two 
$(T_e \approx 0.5-2.0$ MK), i.e., $d_{A,med} \approx 20-100$ Mm.

\subsection{	Secondary CME Acceleration Phase in Heliosphere		} 

While our analysis method based on the Neupert effect places
the CME acceleration region into a low coronal height of
$d_{A,med} \approx 40$ Mm $\approx 0.06\ R_{\odot}$,
this does not exclude secondary acceleration at larger heights.
The height-time plots of the LASCO data reveal secondary
acceleration phases, as well as deceleration phases,
in heights of $r \approx 3-30\ R_{\odot}$, which can be
recognized by their gradual steepening in this height range.
For instance, secondary acceleration are most clearly evident
in the events \#406 (Fig.~3c), \#407 (Fig.~3d), 
\#117 (Fig.~4a), \#468 (Fig.~4b), \#523 (Fig.~4c), and
\#39 (Fig.~4d), which mostly contain CMEs that propagated
over large distances (Fig.~3) or were observed over the
longest durations (Fig.~4). So, there is clear evidence
for secondary acceleration further out in the heliosphere,
but we focus here on the primary acceleration phase only,
which generally is driven by a much higher acceleration rate than
the secondary acceleration phase. It is likely that the
secondary acceleration phase is strongly controlled by
aerodynamic drag effects (Cargill 2004; Vrsnak et al.~2004, 
2013; Aschwanden and Gopalswamy 2019). The main effect
is that CMEs that come out of the coronal primary acceleration
phase faster than the ambient solar wind (which has a typical
speed of $v_B \approx 400$ km s$^{-1}$) (Fig.~8d), will be slowed
down to solar wind speed, and vice versa, CMEs with initially
slow speeds will be accelerated by the solar wind through the
aerodynamic drag force. 

\subsection{	The Neupert Effect 			}

The timing of nonthermal HXR emission provides 
a crucial test for all eruptive flares and CMEs. 
The temporal coincidence of flare HXR emission
with the start of the CME acceleration implies a
causality between the two types of emissions. 
Nonthermal HXRs are believed to be produced by nonthermal
electrons that precipitate from a coronal reconnection
site into the chromosphere according to the thick-target model,
where they are stopped by Coulomb collisions and build up
a high plasma pressure that releases its pressure by
driving upflows and CMEs. While HXR time profiles
$F_{\rm HXR}$ are proportional to the nonthermal electron
flux, the accompanied SXR emission piles up
according to the time integral of the flux, which implies that
its time derivative is proportional to the flux,
as stated in the Neupert effect model. Our test of the
Neupert model requires a coincidence between the
time derivative SXR flux (being the proxy of the
HXR flux) and the extrapolated height-time profile
of the CME motion, which was found to be the case
in $\approx 65\%$ of the (automatically) analyzed events. 

High-temperature plasma ($\gapprox 16.5$ MK) was found to be
more likely than low-temperature plasma to exhibit to
Neupert effect, in which the time derivative of the SXR emission
measure is similar to the light curve of the impulsive
hard X-ray emission for the flare (McTiernan et al.~1999).
A good correlation between occulted hard X-rays (from
RHESSI) and the time derivative of the SXR flux was found
in many flares, which confirms the Neupert effect
in terms of the thin-target bremsstrahlung model, rather
than the thick-target model in non-occulted flares
(Effenberger et al.~2017). 
The Neupert effect has also been tested in UV and
SXR wavelengths, but a two-phase heating model was
required to obtain agreement with observations (Qiu 2021). 

\section{	Conclusions				}

In this study we model the acceleration phase of
CMEs by means of the Neupert effect, which yields 
important physical parameters in the energization of
flare-associated CMEs, such as the peak time $t_0$ of the
CME acceleration phase, the duration $\tau_A$
of the acceleration phase, the height-time profile
$r(t)$, the velocity-time profile $v(t)$, and the
acceleration rate $a(t)$ of propagating CMEs. 
A summary of the conclusions is given in the following. 

\begin{enumerate}
\item{Data analysis and modeling has been applied to a
CME catalog of 576 flare/CME events that includes all 
GOES X- and M-class flares recorded during 2010-2014. 
Combining information from GOES and LASCO/SOHO data 
sets we are able to model the kinematics of the CME 
acceleration and propagation. In this study we attempt 
to connect the time profiles of coronal SXR emission 
in flares with the white-light emission of 
flare-associated CMEs in heliospheric distances 
in a self-consistent way. The observables consist of
CME-observed times $[t_1,...,t_n]$ and 
projected distances $[r_1,...,r_n]$, which are
linearly extrapolated in the intermediate range
$[t_B \le t \le t_1]$ and quadratically
extrapolated during the CME acceleration phase
$[t_A \le t \le t_B]$. 
The missing link is the determination of the
exact timing when a CME is accelerated,
which we derive from the peak time $t_0=(t_B+t_A)/2$
and width (FWHM) $\tau_A=(t_B-t_A)$ in the time derivative 
$dF_{\rm SXR}(t)/dt$ of the GOES SXR flux,
which represent a suitable proxy for the HXR 
flux profiles $F_{\rm HXR}(t)$, according to the
Neupert effect.  The search of the peak time of the time derivative
is limited to the flare SXR rise time interval $\tau_{\rm rise}$.}

\item{We find the following medium values and parameter
ranges in our statistical sample of $n \lapprox 576$ events,
shown in form of histograms in Fig.~(8): 
SXR rise time $\tau_{\rm rise}=9.0$ min $(1.0 \le \tau_{\rm rise} \le 311.0)$ min;
acceleration duration $\tau_A=3.0$ min $(1.2 \le \tau_A \le 44.5)$ min;
acceleration distance $d_A=39$ min $(2.8 \le d_A \le 1063$ Mm); and
acceleration rate $a_A=2.0$ km s$^{-2}$ $(0.1 \le a_A \le 13.5)$ Mm.
Note that the acceleration time duration is about three times smaller
than the SXR rise time, i.e., $\tau_A/\tau_{\rm rise} \approx 0.3$.}  

\item{In order to investigate possible physical scaling laws we
plot the cross-correlations of 6 parameter pairs (Fig.~9). The
strongest correlations are found for the CME acceleration distance
$d_A$ versus the CME acceleration duration $\tau_A$, with CCC=0.70 (Fig.~9b),
and for the CME velocity $v_B$ versus the CME acceleration distance $d_A$,
with CCC=0.73 (Fig.~9d). The CME distance from Sun center is defined here
by $d_A=r_B-r_A$, which corresponds to the distance traveled through
the acceleration region (in radial direction). If the CME velocities
have a relatively small variation, we can understand the first correlation
$v_B=(r_B-r_A)/(t_B-t_A) \propto d_A/\tau_A$, from which the second
correlation follows also. 
A marginal correlation (CCC=0.41, Fig.~9a) is found between the
CME velocity and the GOES flux, which is similar to 
an earlier study (Moon et al.~2002).
However, only a weak correlation
(CCC=0.31) is found between the CME acceleration time
and the SXR rise time, which indicates that
substantial parts of the SXR emission do not
produce HXR emission.}

\item{The CME propagation distance $d_A=(r_B-r_A)$ 
during the time interval $[t_B, t_A]$, marks the 
vertical extent of the acceleration region, which is
is found to have a median value of $d_A \approx 40$ Mm and
matches the hydrostatic scale height $\lambda(T)$ of the
Quiet-Sun corona for a mean temperature of $T \approx 0.8$ MK.
This result implies that CME acceleration occurs in the
lowest scale height of the hydrostatic solar corona
at $r \lapprox 1.1 \ R_{\odot}$, while secondary 
acceleration possibly observed in the heliospheric path
of $r \approx 3-30\ R_{\odot}$ are weaker than
the primary acceleration rate in the lower corona.}

\end{enumerate}

The Neupert effect serves as a suitable proxy for HXR 
$F_{\rm HXR}(t)$ and can simply be obtained from the time
derivative of the SXR flux $F_{\rm SXR}$. The
usage of the Neupert effect is particularly useful in times
when no (solar-dedicated) HXR detectors are available,
such as presently, after the demise of RHESSI in 2018.
However, it remains to be shown how accurately the Neupert
proxy represents HXR emission (Dennis and Zarro 1993).
Nevertheless, as this study shows, the Neupert effect
helps enormously to bridge the coronal to the heliospheric
part of propagating CMEs, by using a fully automated data analysis
code. Future work may include STEREO data (Temmer et al.~2010), 
which has a higher temporal cadence and spatial coverage 
in the lower corona ($r \lapprox 3\ R_{\odot}$) that are 
occulted in coronagraphs like LASCO/SOHO. Coverage in this
distance range is crucial to study acceleration and
deceleration by the solar wind, by including 
aerodynamic drag effects (Cargill 2004;
Vrsnak et al.~2004, 2013; Aschwanden and Gopalswamy 2019).

\bigskip
{\sl Acknowledgements:}
Part of the work was supported by NASA contracts 
NNG04EA00C and NNG09FA40C.  

\clearpage

%%%%%%%%%%%%%%%%%%%%%%%%%%% REFERENCES &&&&&&&&&&&&&&&&&&&&&&&&&&&&&&&
\section*{ References }  

\def\ref#1{\par\noindent\hangindent1cm {#1}} 

\ref{Aschwanden, M.J. and Alexander, D. 2001, SolPhys 204, 91.}
 	% Flare Plasma Cooling from 30 MK down to 1 MK modeled 
	% from Yohkoh, GOES, and TRACE observations during the 
	% Bastille-Day Event (2000 July 14)
\ref{Aschwanden, M.J. 2004, {\sl Physics of the Solar Corona - 
	An Introduction (1st Edition)}, Springer: New York}
\ref{Aschwanden, M.J. 2016, ApJ 831, 105, (Paper IV)}
	% Global energetics of solar flares: IV. Coronal Mass 
	% Ejection Energetics
\ref{Aschwanden, M.J. 2017, ApJ 847:27, (Paper VI)}
 	% Global energetics of solar flares: VI. Refined energetics 
	% of coronal mass ejection Energetics
\ref{Aschwanden, M.J. and Gopalswamy, N. 2019, ApJ 877:149, (Paper VII)} 
 	% Global energetics of solar flares: VII. Aerodynamic drag in 
	% coronal mass ejections
\ref{Burkepile,J.T., Hundhausen, A.J., Stanger, A.L., St. Cyr, O.C., 
	and Seiden, J.A. 2004, JGR (Space Physics), 3103}
 	% Role of projection effects on solar coronal mass ejection 
	% properties: 1. A study of CMEs associated with limb activity
\ref{Cargill, P.J. 2004, SoPh 221, 135}
 	% On the Aerodynamic Drag Force Acting on Interplanetary Coronal 
	% Mass Ejections
\ref{Dennis, B.R. and Zarro, D.M. 1993, SoPh 146, 177}
 	% The Neupert effect: what can it tell us about the impulsive 
	% and gradual phases of solar flares
\ref{Effenberger, F., Rubio da Costa, F., Oka, M., Saint-Hilaire P.,
	Liu, W., Petrosian, V., Glesener, L., and Krucker, S. 2017, ApJ 835:124}
	% Hard X-ray emission from partially occulted solar flares: RHESSI 
	% observations in two solar cycles 
\ref{Gopalswamy, N., Yashiro, S., Kaiser, M. L., Howard, R. A., and
	Bougeret, J. 2001, ApJ 548, L91}
 	% Radio Signatures of Coronal Mass Ejection Interaction: 
	% Coronal Mass Ejection Cannibalism ?
\ref{Gopalswamy,N., Yashiro,S., Michalek,G., Stenborg,G., Vourlidas,A., Freeland,S., 
	and Howard,R. 2009, Earth, Moon, and Planets 104, 295}
 	% The SOHO/LASCO CME catalog
\ref{Gopalswamy,N., Yashiro, S., Michalek, G., Xie, H., Maekelae, P., Vourlidas, A., 
	and Howard, R.A. 2010, Sun and Geosphere, 5, 7}
 	% A Catalog of Halo Coronal Mass Ejections from SOHO
\ref{Howard,T.A. and Harrison, R.A. 2013, SolPhys 285, 269}
	% Stealth Coronal Mass Ejections: A Perspective
\ref{Hudson, H.S. 1991, BAAS 23, 1064}
 	% Differential Emission-Measure variations and the Neupert Effect
\ref{Maricic,D., Vrsnak, B., Stanger, A. L., and Veronig, A. 2004, SP 225, 337}
	% Coronal Mass Ejection of 15 May 2001: I. Evolution of Morphological 
	% Features of the Eruption
\ref{Maricic,D., Vrsnak, B., Stanger, A. L., Veronig, A. M., 
	  Temmer, M., and Rosa, D. 2007, SolPhys 241, 99}
 	% Acceleration Phase of Coronal Mass Ejections: 
	% II. Synchronization of the Energy Release in the 
	% Associated Flare
\ref{Moon, Y.J., Choe,G.S., Wang,H., Park,Y.D., Gopalswamy,N., Yang,G., 
	and Yashiro,S. 2002, ApJ 581, 694}
 	% A statistical study of two classes of coronal mass ejections
\ref{Moon,Y.J., Choe,G.S., Wang,H., Park,Y.D., and Cheng,C.Z.
 	2003, JKAS 36, 61}
 	% Relationship between CME kinematics and flare strength
\ref{Kahler,S.W. and Cliver,E.W. 1988, SolPh 115, 385}
 	% Solar cycle variation of long duration 10.7 cm and 
	% soft X-ray bursts
\ref{McTiernan, J.M., Fisher, G.H., and Li, P. 1999. ApJ 514:472}
	% The solar flare soft X-ray differential emission measure
	% and the Neupert effect at different temperatures
\ref{Neupert, W.M. 1968, ApJ 153, L59}
 	% Comparison of solar X-ray line emission with microwave 
	% emission during flares
\ref{Qiu,J., Wang, H., Cheng, C.Z., and Gary, D. 2004, ApJ 604:900}
	% Magnetic reconnection and mass acceleration in flare coronal
	% mass ejection events
\ref{Qiu,J. 2021, ApJ 909:99}
	% The Neupert effect of flare ultraviolet and soft X-ray emissions
\ref{Shanmugaraju,A., Moon, Y.-J., Dryer, M., and Umapathy, S.
 	2003, SolPhys 215, 185}
 	% On the kinematic evolution of flare-associated CMEs
\ref{Temmer, M., Veronig, A.M., Vrsnak, B., Rybak, J., Gomory,P., 
	Stoiser, S., Maricic,D. 2008, ApJ 673, L95}
 	% Acceleration in fast halo CMEs and synchronized flare HXR bursts
\ref{Temmer, M., Veronig, A.M., Kontar, E.P., Krucker, S., Vrsnak, B.
 	2010, ApJ 712, 1410}
 	% Combined STEREO/RHESSI study of CME acceleration and particle 
	% acceleration in solar flares
\ref{Veronig, A.M., Vrsnak, B., Dennis, B.R., Temmer, M., Hanslmeier, A., 
	and Magdalenic, J. 2002, A\&A 392, 699}
 	% Investigation of the Neupert effect in solar flares. 
	% I. Statistical properties and the evaporation model
\ref{Veronig,A.M., Brown,J.C., Dennis,B.R., Schwartz,R.A., Sui,L., and Tolbert,A.K.
	2005 ApJ 621, 482}
        % Physics of the Neupert Effect: Estimates of the effects of source energy 
        % and mass transport, and geometry, using RHESSI and GOES data
\ref{Vrsnak,B., Ruzdjak, D., Sudar, D., and Gopalswamy, N.
 	2004, AA 423, 717}
 	% Kinematics of coronal mass ejections between 2 and 30 
	% solar radii. What can be learned about forces governing 
	% the eruption?
\ref{Vrsnak,B., Sudar, D., and Ruzdjak, D. 2005, AA 435, 1149}
 	% The CME-flare relationship: Are there really two types 
	% of CMEs?
\ref{Vrsnak,B., Maricic, D., Stanger, A. L., Veronig, A. M., 
	Temmer, M., and Rosa, D. 2007, SolPhys 241, 85}
 	% Acceleration Phase of Coronal Mass Ejections: 
	% I. Temporal and Spatial Scales
\ref{Vrsnak, B., Zic, T., Vrbanec, D., Temmer, M., Rollett, T., et al.
 	2013, SoPh, 285, 295}
 	% Propagation of Interplanetary Coronal Mass Ejections: 
	% The Drag-Based velocity Model
\ref{Zhang, Jie, Dere, K.P., Howard, R.A., Kundu, M.R., and White, S.M.
 	2001, ApJ 559, 452}
 	% On the Temporal Relationship between Coronal Mass Ejections 
	% and Flares
\ref{Zhang, Jie, Dere,K.P., Howard,R.A., and Vourlidas,A.
 	2004, ApJ 604, 420}
 	% A study of the kinematic evolution of coronal mass ejections
\ref{Zhang,J., and Dere, K. P.  2006, ApJ 649, 1100}
 	% A Statistical Study of Main and Residual Accelerations 
	% of Coronal Mass Ejections

\clearpage %%%@@@

%%%%%%%%%%%%%%%%%%%%%%%%%%% REF2 %%%%%%%%%%%%%%%%%%%%%%%%%%%%%%%%

%%%%%%%%%%%%%%%%%%%%%%%%%%% TABLE 1 &&&&&&&&&&&&&&&&&&&&&&&&&&&&&&&&&
\begin{table}
\begin{center}
\footnotesize 		%\tiny \scriptsize footnotesize small normal
\caption{Measurements of timing $[t_A, t_B, t_1, t_n]$, distances from Sun center $[r_A, r_B, r_1, r_n]$, velocity $[v_B]$,
	and acceleration rate $[a_A]$ of CMEs and solar flares based on predictions by the Neupert effect,
	 for 24 selected flares (Table 1 and Figures 1-6).} 
\medskip
\begin{tabular}{rrrrrrrrrrrrrrr}
\hline
Fig &  Nr & GOES &  HELPOS &        DATEOBS &   $t_A$ &   $t_B$ &   $t_1$ &   $t_n$ &     $r_A$ &    $r_B$ &   $r_1$ &   $r_n$ &  $v_B$ &  $a_A$ \\      
    &     &      &         & yyyymmdd hhmm  &    [hr] &    [hr] &    [hr] &    [hr] & [$R_{\odot}$] & [$R_{\odot}$] & [$R_{\odot}$] & [$R_{\odot}$] & [km/s] & [km/s$^2$] \\
\hline
\hline
1a &  61 & X6.9 & N20W69 & 2011-08-09 07:48 &   8.037 &   8.076 &   8.202 &  10.702 &   0.950 &   1.139 &   3.010 &  24.520 &   1830 &  12.778 \\
1b & 147 & X5.4 & N18E31 & 2012-03-07 00:02 &   0.256 &   0.352 &   0.402 &   2.301 &   0.586 &   1.285 &   2.360 &  28.690 &   2813 &   8.133 \\
1c & 437 & X4.9 & S12E82 & 2014-02-25 00:39 &   0.740 &   0.792 &   1.507 &   2.702 &   0.992 &   1.317 &  11.170 &  24.590 &   2426 &  13.025 \\
1d & 344 & X3.3 & S08E42 & 2013-11-05 22:07 &  22.181 &  22.212 &  22.601 &  28.502 &   0.679 &   0.733 &   3.330 &  20.730 &    674 &   5.987 \\
   &     &      &        &                  &         &         &         &         &         &         &         &         &        &         \\
2a &  35 & C7.3 & N10W05 & 2011-03-09 10:35 &  11.016 &  11.080 &  12.202 &  14.702 &   0.194 &   0.239 &   3.040 &   7.100 &    266 &   1.143 \\
2b & 221 & M1.0 & S20W23 & 2012-07-14 04:51 &   4.895 &   4.944 &   5.001 &   7.101 &   0.507 &   0.550 &   4.740 &   8.090 &    333 &   1.874 \\
2c & 426 & M1.0 & S12W13 & 2014-02-13 08:05 &  10.518 &  10.561 &  11.601 &  14.001 &   0.304 &   0.326 &   2.810 &   6.210 &    200 &   1.305 \\
2d & 293 & M1.0 & N12E42 & 2013-05-31 19:52 &  19.928 &  19.987 &  20.601 &  23.902 &   0.691 &   0.768 &   2.870 &   9.970 &    505 &   2.368 \\
   &     &      &        &                  &         &         &         &         &         &         &         &         &        &         \\
3a & 146 & M1.0 & N18E32 & 2012-03-06 22:49 &   0.256 &   0.352 &   0.402 &   2.301 &   0.598 &   1.297 &   2.360 &  28.690 &   2813 &   8.133 \\
3b & 147 & X5.4 & N18E31 & 2012-03-07 00:02 &   0.256 &   0.352 &   0.402 &   2.301 &   0.586 &   1.285 &   2.360 &  28.690 &   2813 &   8.133 \\
3c & 406 & M1.0 & S11E13 & 2014-02-02 16:24 &  16.444 &  16.482 &  17.401 &  28.702 &   0.293 &   0.312 &   2.440 &  28.530 &    195 &   1.452 \\
3d & 407 & M3.1 & S11E13 & 2014-02-02 18:05 &  16.444 &  16.482 &  17.401 &  28.702 &   0.293 &   0.312 &   2.440 &  28.530 &    195 &   1.452 \\
   &     &      &        &                  &         &         &         &         &         &         &         &         &        &         \\
4a & 117 & M1.9 & S19E36 & 2011-11-15 12:30 &  12.636 &  12.704 &  12.801 &  33.701 &   0.652 &   0.666 &   2.950 &  20.310 &     80 &   0.328 \\
4b & 468 & M1.0 & S10W57 & 2014-05-06 22:01 &  22.099 &  22.141 &  22.294 &  39.702 &   0.847 &   0.854 &   3.090 &  15.200 &     70 &   0.471 \\
4c & 523 & M3.9 & S13E71 & 2014-10-20 09:00 &   3.903 &   3.941 &   5.401 &  20.301 &   0.952 &   0.956 &   3.940 &  16.220 &     41 &   0.299 \\
4d &  39 & M1.3 & N07W41 & 2011-03-12 04:33 &  18.642 &  18.676 &  20.202 &  34.701 &   0.664 &   0.673 &   2.500 &  22.910 &    104 &   0.852 \\
   &     &      &        &                  &         &         &         &         &         &         &         &         &        &         \\
5a & 209 & X1.1 & S13W59 & 2012-07-06 23:01 &  23.094 &  23.133 &  23.402 &  25.502 &   0.870 &   1.057 &   5.030 &  25.050 &   1873 &  13.459 \\
5b & 287 & X3.2 & N08E77 & 2013-05-14 01:00 &   1.102 &   1.169 &   1.431 &   2.701 &   0.976 &   1.471 &   5.840 &  23.030 &   2851 &  11.806 \\
5c & 437 & X4.9 & S12E82 & 2014-02-25 00:39 &   0.740 &   0.792 &   1.507 &   2.702 &   0.992 &   1.317 &  11.170 &  24.590 &   2426 &  13.025 \\
5d & 131 & M8.7 & N30W21 & 2012-01-23 03:38 &   3.740 &   3.910 &   4.001 &   6.101 &   0.597 &   1.564 &   3.380 &  27.640 &   2201 &   3.596 \\
   &     &      &        &                  &         &         &         &         &         &         &         &         &        &         \\
6a & 217 & M1.1 & S17E38 & 2012-07-09 23:03 &  19.813 &  19.866 &  21.401 &  27.501 &   0.664 &   0.686 &   2.600 &  10.170 &    161 &   0.861 \\
6b & 124 & M1.2 & S25E67 & 2011-12-30 03:03 &  -0.289 &  -0.245 &   1.431 &   9.901 &   0.948 &   0.974 &   3.330 &  15.240 &    228 &   1.469 \\
6c & 306 & M1.3 & S21W22 & 2013-10-15 23:31 &  20.132 &  20.178 &  21.418 &  27.702 &   0.506 &   0.522 &   2.830 &  10.420 &    130 &   0.787 \\
6d &  64 & M1.2 & N18W87 & 2011-09-05 07:27 &   2.642 &   2.668 &   3.401 &  10.102 &   1.000 &   1.014 &   2.550 &  12.040 &    212 &   2.258 \\
\hline
\end{tabular}
\end{center}
\end{table}

%%%%%%%%%%%%%%%%%%%%%%%%%%% TABLE 2 &&&&&&&&&&&&&&&&&&&&&&&&&&&&&&&&&
\begin{table}
\begin{center}
\footnotesize 		%\tiny \scriptsize footnotesize small normal
\caption{Measurements of timing $[t_A, t_B, t_1, t_n]$, altitudes $[r_A, r_B, r_1, r_n]$, velocity $[v_B]$,
	and acceleration rate $[a_A]$ of CME kinematics. 
	The full table of 576 events is available as a machine-readable file.}
\medskip
\begin{tabular}{rrrrrrrrrrrrrr}
  Nr & GOES & HELPOS &  DATEOBS         &   $t_A$ &   $t_B$ &   $t_1$ &   $t_n$ &     $r_A$ &    $r_B$ &   $r_1$ &   $r_n$ &  $v_B$ &  $a_A$ \\      
    &       &        &  yyyymmdd hhmm   &    [hr] &    [hr] &    [hr] &    [hr] & [$R_{\odot}$] & [$R_{\odot}$] & [$R_{\odot}$] & [$R_{\odot}$] & [km/s] & [km/$s^2$] \\
\hline
\hline
   1 & M2.0 & N23W47 & 2010-06-12 00:30 &   0.937 &   0.973 &   1.528 &   9.719 &   0.791 &   0.842 &   3.090 &  23.090 &    546 &   4.271 \\
   2 & M1.0 & S24W82 & 2010-06-13 05:30 &   5.592 &   5.638 &   5.901 &  16.301 &   0.997 &   1.021 &   3.280 &  19.470 &    205 &   1.239 \\
   3 & M1.0 & N13E34 & 2010-08-07 17:55 &  18.161 &  18.359 &  18.602 &  23.702 &   0.593 &   1.106 &   3.560 &  26.870 &   1005 &   1.417 \\
   4 & M2.9 & S18W26 & 2010-10-16 19:07 &  19.170 &  19.205 &  20.202 &  21.819 &   0.524 &   0.549 &   2.580 &   5.510 &    278 &   2.230 \\
   5 & M1.6 & S20E85 & 2010-11-04 23:30 &  -0.101 &  -0.041 &   0.201 &   3.502 &   0.999 &   1.030 &   4.430 &   8.420 &    198 &   0.923 \\
   6 & M1.0 & S20E75 & 2010-11-05 12:43 &  -0.100 &  -0.041 &   1.430 &   4.302 &   0.977 &   1.002 &   4.950 &   9.350 &    161 &   0.750 \\
   7 & M5.4 & S20E58 & 2010-11-06 15:27 &  15.547 &  15.604 &  16.202 &  20.501 &   0.878 &   0.897 &   2.450 &   6.450 &    135 &   0.661 \\
   8 & M1.3 & N16W88 & 2011-01-28 00:44 &   0.942 &   0.980 &   1.429 &   5.301 &   1.000 &   1.077 &   3.000 &  15.150 &    784 &   5.718 \\
   9 & M1.9 & N16W70 & 2011-02-09 01:23 &   0.178 &   0.220 &   0.401 &   2.202 &   0.950 &   0.969 &   2.790 &   4.640 &    173 &   1.160 \\
  10 & M6.6 & S21E04 & 2011-02-13 17:28 &  17.556 &  17.604 &  18.601 &  22.102 &   0.365 &   0.399 &   2.680 &   9.350 &    278 &   1.600 \\
 ... &  ... &  ...   & ...              &   ...   &   ...   &   ...   &     ... &     ... &     ... &     ... &     ... &    ... &     ... \\

\hline
\end{tabular}
\medskip 
\end{center}
\end{table}

%_______________________________TABLE 3_________________________
\begin{table}
\begin{center}
\small 		%\tiny \scriptsize footnotesize small normal
\caption{Statistics of analyzed events.}
\medskip
\begin{tabular}{lrr}
\hline
\hline
Total number of analyzed events  &	    576 &      100\% 	\\
Events with CME preceding flares &          131 &       23\% 	\\
Ambiguous flare/CME association  &           84 &       15\% 	\\
Consistent with Neupert model    &          373 &       65\%    \\
\hline
\end{tabular}
\medskip 
\end{center}
\end{table}

\clearpage

%%%%%%%%%%%%%%%%%%%%%%%%%%%%%%%% FIGURES %%%%%%%%%%%%%%%%%%%%%%%%%%%%%%%%

\begin{figure}		
\centerline{\includegraphics[width=0.9\textwidth]{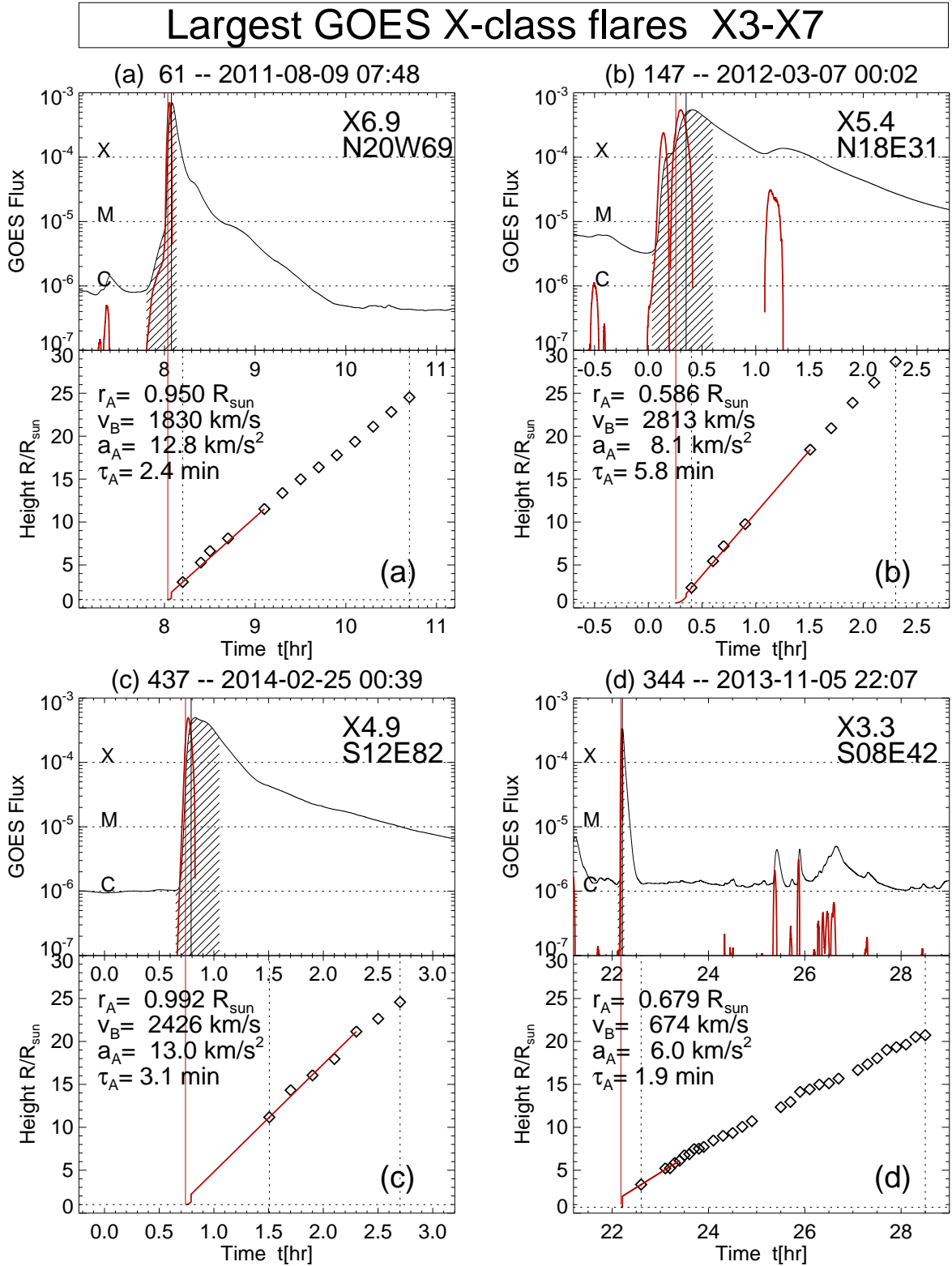}}
\caption{Selection of 4 CME events associated with the largest GOES flares.
The GOES 1-8 \ang\ flux is shown with a black curve. The duration
of the flare, encompassed by the start time and end time as defined
by the NOAA flare classification, is indicated with a dashed area.
The time derivative is indicated with red color, and the peak, as 
defined by the Neupert effect, is marked with a red vertical line.
The height-time profile of the LASCO-observed CME is marked with
black diamonds, and the first and last observed times with vertical
dotted lines.}
\end{figure}

\begin{figure}		
\centerline{\includegraphics[width=1.0\textwidth]{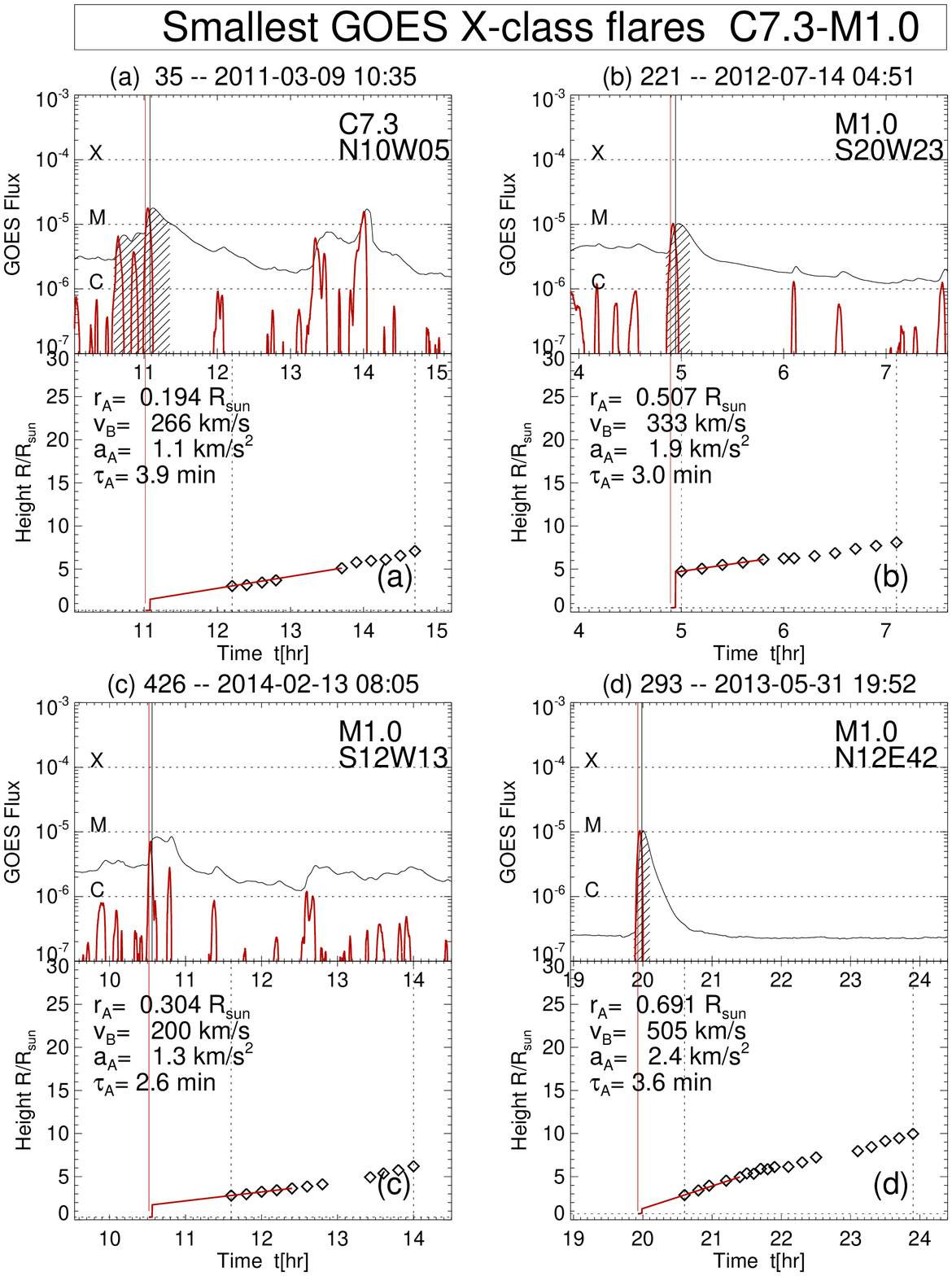}}
\caption{Selection of 4 CME events with smallest GOES flares,
otherwise similar to representation in Fig.~1.}
\end{figure}

\begin{figure}		
\centerline{\includegraphics[width=1.0\textwidth]{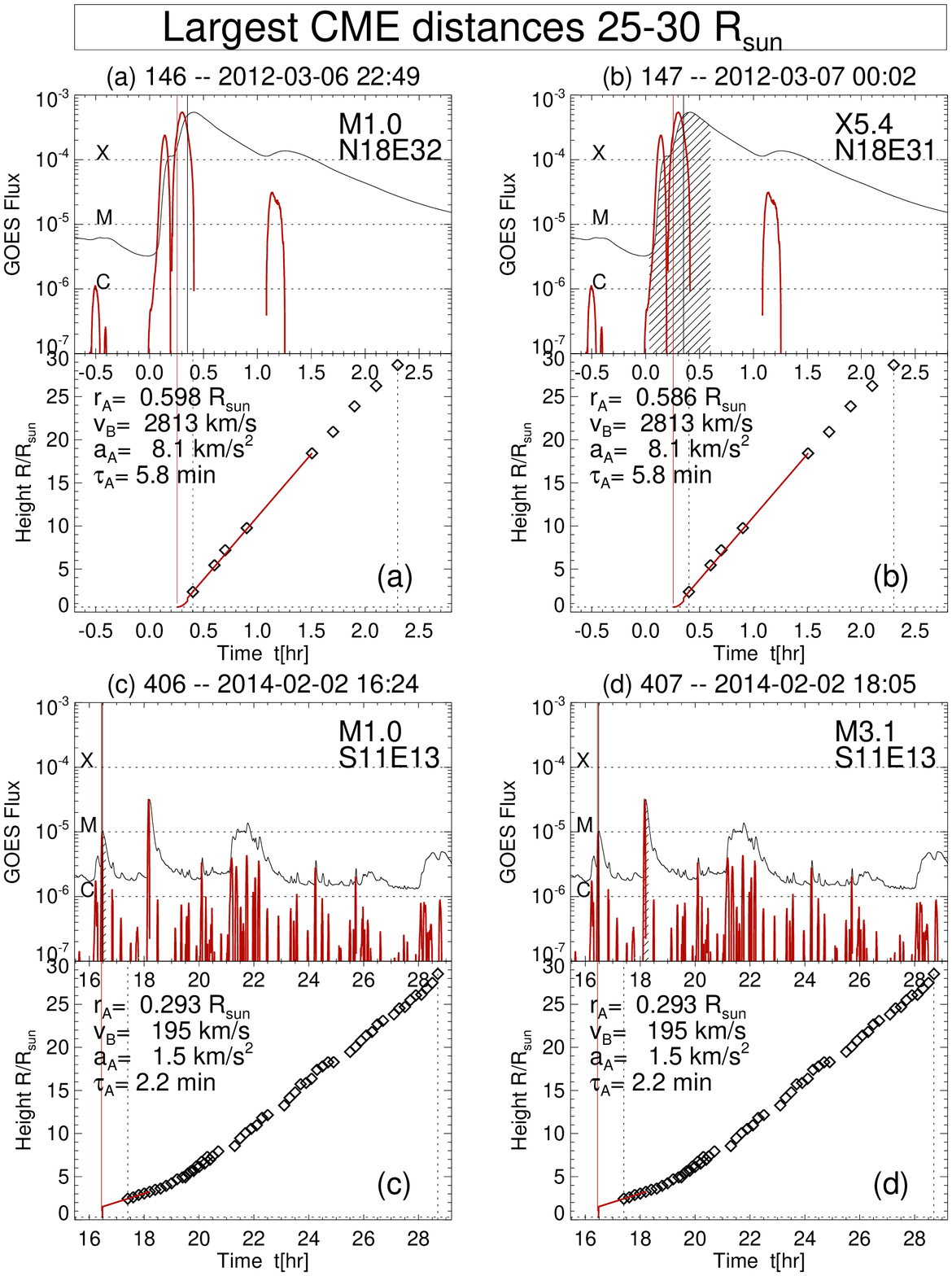}}
\caption{Selection of events with largest CME detection distances,
otherwise similar to representation in Fig.~1.}
\end{figure}

\begin{figure}		
\centerline{\includegraphics[width=1.0\textwidth]{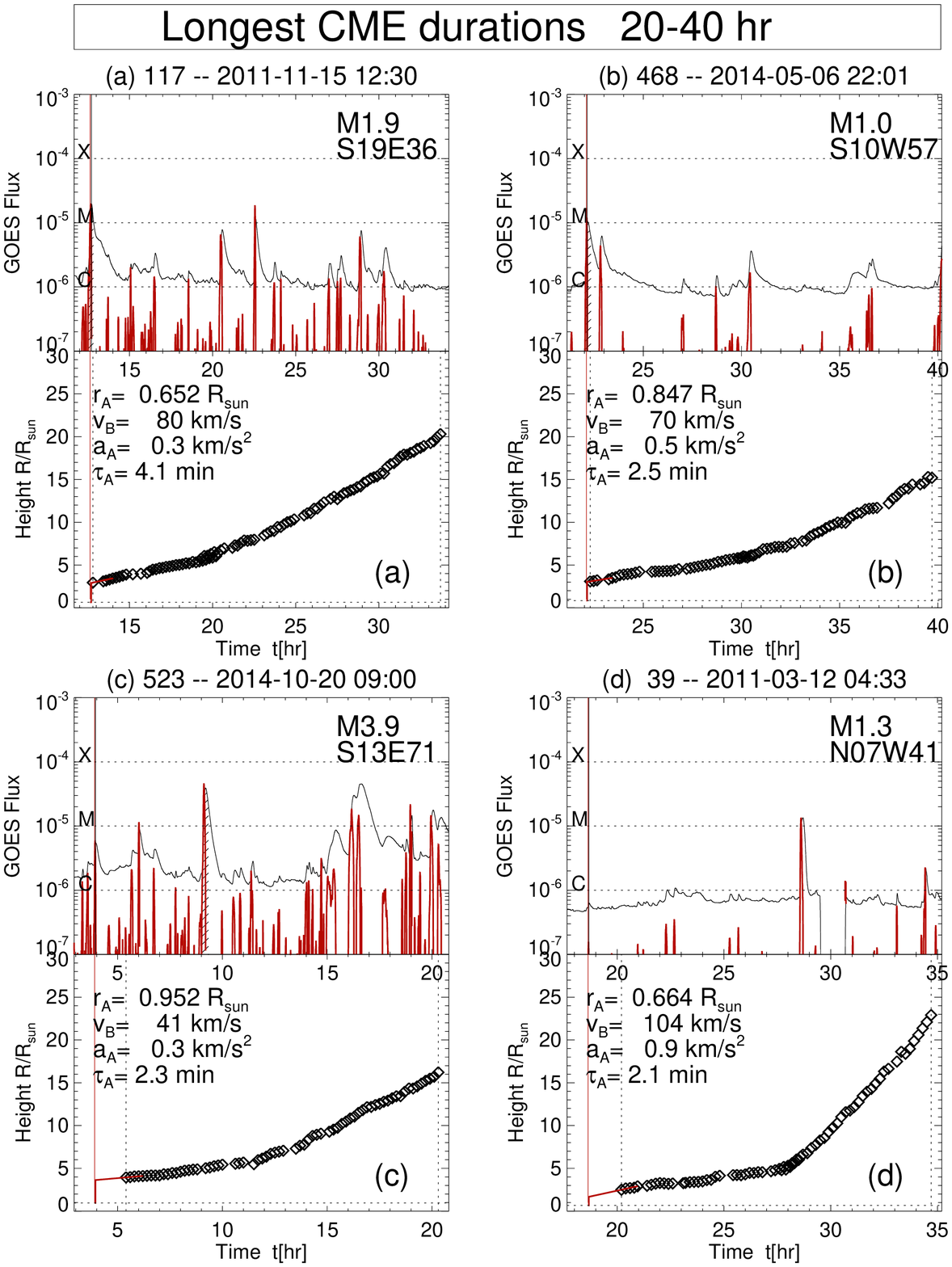}}
\caption{Selection of events with largest CME detection times,
otherwise similar to representation in Fig.~1.}
\end{figure}

\begin{figure}		
\centerline{\includegraphics[width=1.0\textwidth]{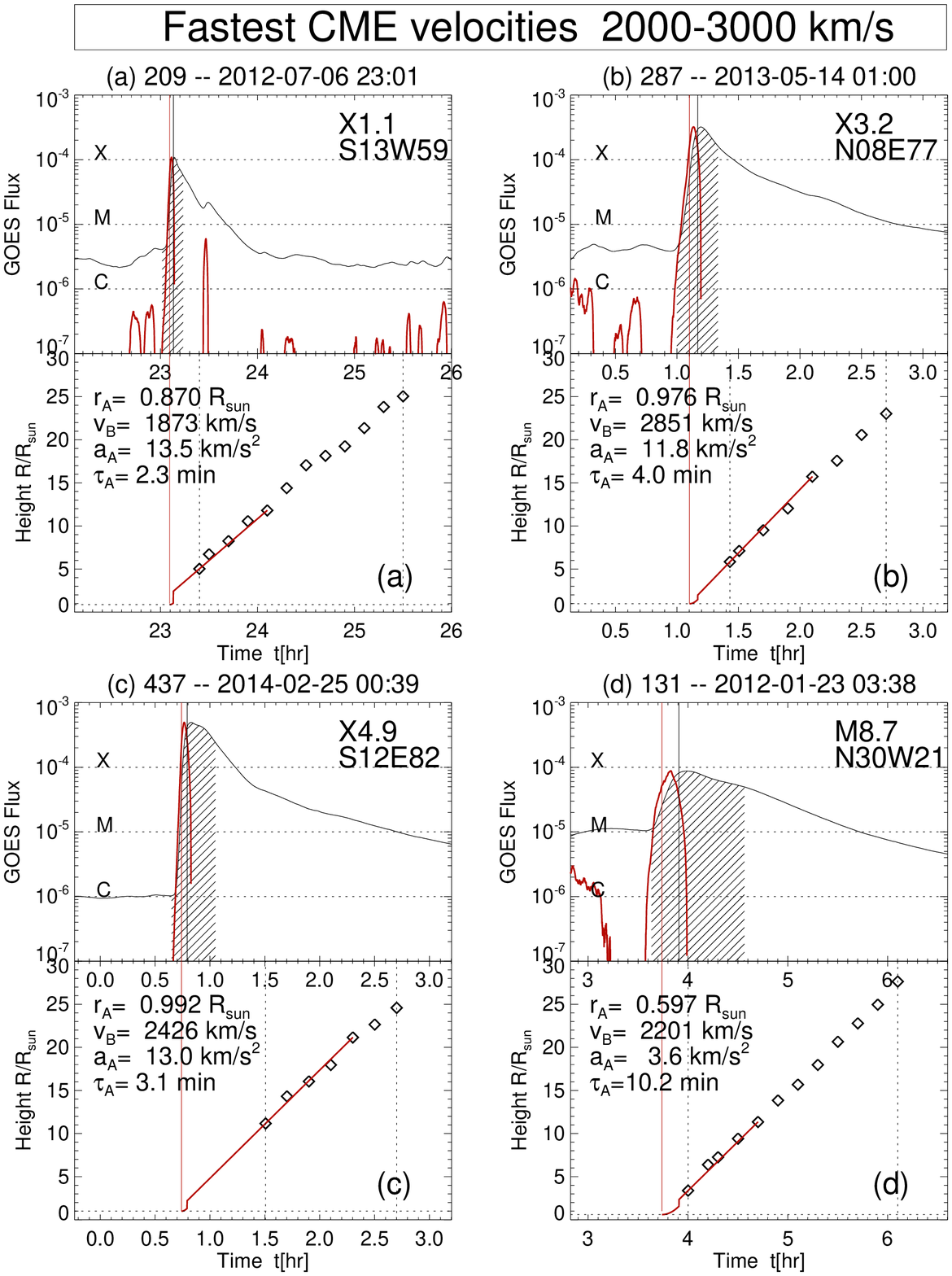}}
\caption{Selection of events with largest CME velocities,
otherwise similar to representation in Fig.~1.}
\end{figure}

\begin{figure}		
\centerline{\includegraphics[width=1.0\textwidth]{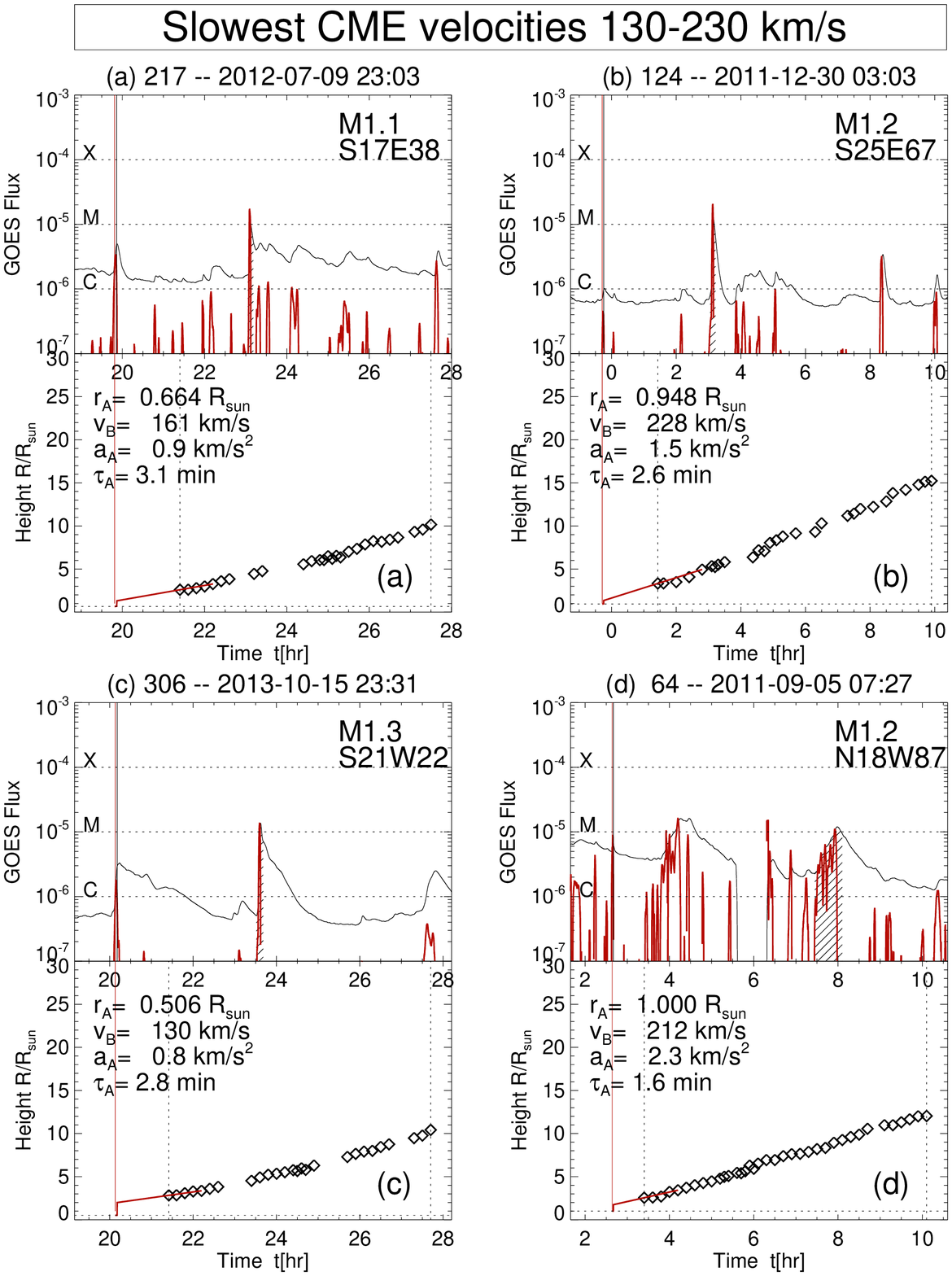}}
\caption{Selection of events with slowest CME velocities,
otherwise similar to representation in Fig.~1.}
\end{figure}

\begin{figure}		
\centerline{\includegraphics[width=0.9\textwidth]{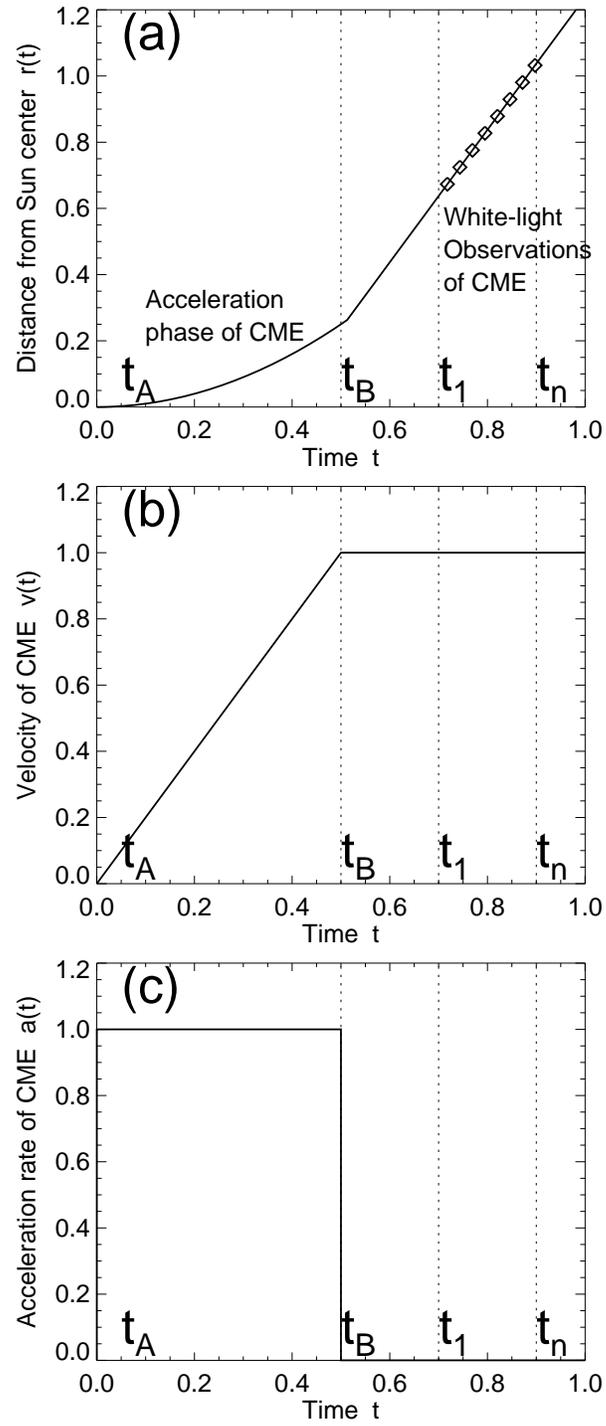}}
\caption{The three panels show the analytical functions that
describe the height-time profile $r(t)$ (a), the velocity
profile $v(t)$ (b), and the acceleration rate $a(t)$ (c).
The white-light observations are indicated with diamonds
($t_1 \le t \le t_n$), and the acceleration occurs during
the time interval ($t_A \le t \le t_B$).}
\end{figure}

\begin{figure}		
\centerline{\includegraphics[width=0.9\textwidth]{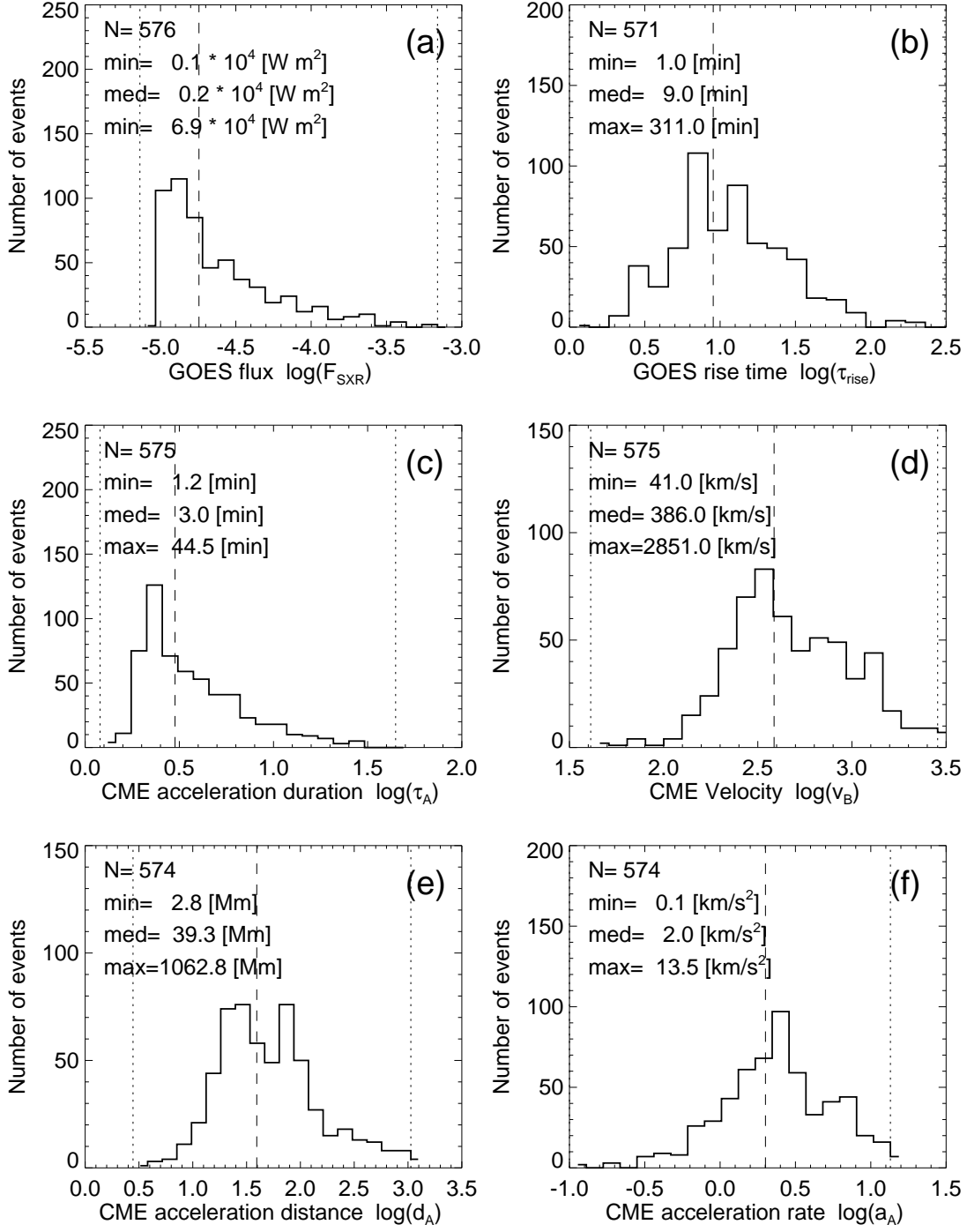}}
\caption{Histograms of the (logarithmic) soft X-ray flux $F_{\rm SXR}$ (a),
the GOES flare rise time $\tau_{\bf rise}$ (b),
the CME acceleration duration $\tau_A=(t_B-t_A)$ (c), 
the CME velocity $v_B=(r_B-r_A)$ (d),
the CME acceleration distance $d_A=(r_B-r_A)$ (e),
and the CME acceleration rate $a_A$ (f). The median values
of the distributions are indicated with dashed lines,
and the minimum and maximum values with dotted lines.}
\end{figure}

\begin{figure}		
\centerline{\includegraphics[width=1.0\textwidth]{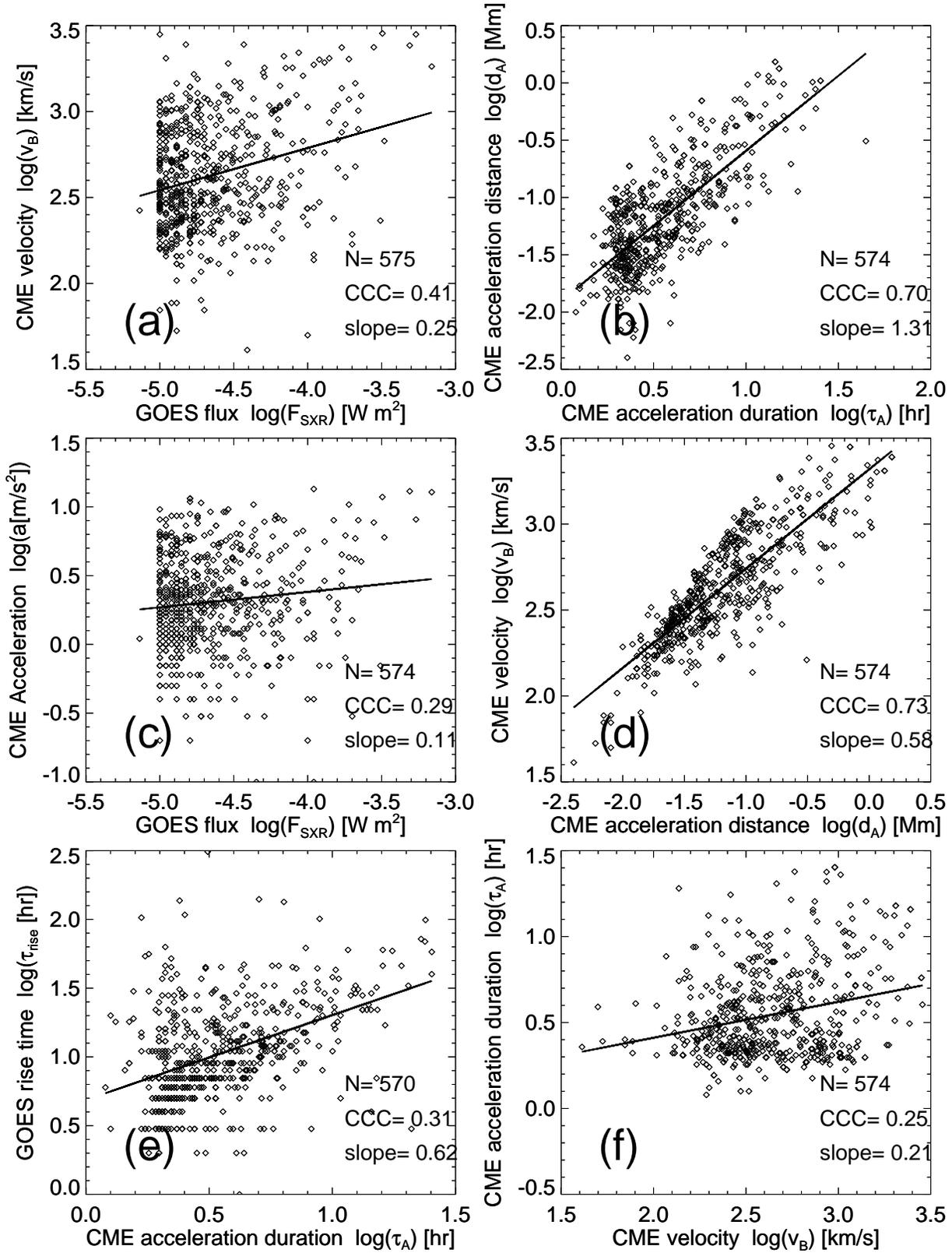}}
\caption{Correlations between the CME velocity and the GOES flux (a), 
the CME acceleration duration and distance (b),
the CME acceleration and the GOES flux (c), 
the CME acceleration distance and velocity (d), and
the CME acceleration time and GOES rise time (e),
the CME acceleration velocity and duration (f). The slopes 
of the linear regression fits and the cross-correlation
coefficients are listed in each panel.}
\end{figure}

\end{document}